\documentstyle [12pt]{article}

\begin{document}


\begin{flushright}
TCD--5--93 \\
August 1993
\end{flushright}

\vspace{8mm}

\begin{center}

{\Large\bf Scenario for Seeding a Singularity  \\
\vspace{3ex}
in $d = 2$ String Black Hole with Tachyon }  \\
\vspace{12mm}
{\large S. Kalyana Rama}

\vspace{4mm}
School of Mathematics, Trinity College, Dublin 2, Ireland. \\
\vspace{1ex}
email: kalyan@maths.tcd.ie   \\
\end{center}

\vspace{4mm}

\vspace{4mm}

\begin{quote}
ABSTRACT.  The $d = 2$ string admits a black hole solution and also
a singular solution when tachyon back reaction is included.
It is of importance to know if the former solution can evolve into
a later one. An explicit solution describing this process is
difficult to obtain. We present here a scenario in which such an
evolution is very likely to occur. In essence, it takes place when
a derivative discontinuity is seeded in the dilaton field
by an incident tachyon pulse.
An application of this scenario to $1 + 1$ dimensional
toy models suggests that a black hole can evolve into
a massive remnant, strengthening its candidacy for the end state of
a black hole.

\end{quote}

\newpage

\vspace{4ex}

\noindent I.
The discovery of a black hole solution in $d = 2$ string \cite{rest}
and Hawking radiation in the string inspired $1 + 1$ dimensional
toy models \cite{cghs}
has generated a great interest in understanding the
black hole dynamics in two dimensions. Models have been constructed
incorporating quantum gravity effects and the back reaction of Hawking
radiation to various degrees.
However the underlying black hole physics is not
completely understood even in this simpler context.

One thing of interest would be the end state of a black hole
when it has stopped emitting Hawking radiation ---
dark black hole or DBH for short. There are various
proposals for DBH but none of them is free of problems
\cite{cghs}-\cite{info}. It is natural to expect that
the static solutions of the above models, besides the black hole ones,
may represent such end states and to look for them.

Indeed such static solutions have been
discovered in $d = 2$ strings in \cite{r}.
In $1 + 1$ dimensional toy models they were discovered
in \cite{suss,num} and shown not to Hawking radiate.
They were also analysed as representatives of massive remnants, one of the
candidates for DBH. These solutions have singular horizons and
are not black holes in the usual sense.

Mere existence of a static solution is not a proof that it is a
DBH since a physical
black hole might not dynamically evolve into it. To understand
the dynamical evolution of black holes, one needs to
solve the relevent (non linear) equations for any arbitrary
incident matter wave. But this is a difficult task.

As explained
in the review papers in \cite{cghs}, the $d = 2$ string $\beta$-function
equations also describe the $1 + 1$ dimensional toy models which include
the back reaction effects of Hawking radiation (see \cite{cghs} for
further details). Thus, the static solutions of $d = 2$ string would
describe the end state of black holes in these toy models, and hence,
the study of dynamical evolution of black hole in $d = 2$ string would
be helpful in understanding the black hole physics.

In this paper, we describe a scenario in the context of $d = 2$
string where the fields originally in the black hole background
evolve nonlinearly into divergent configuration, possibly the
ones described in \cite{r}. The impetus for this evolution
can be provided by a localised tachyon pulse peaked sharply enough.
This pulse (if it is a $\delta$-function)
seeds a cusp, {\em i.e.}\  a derivative discontinuity,
in the dilaton field
configuration. Analysing the equations
in the nearby region, we find that
both the cusp and the dilaton field grow as they evolve towards the
horizon, ending very likely in a singular
configuration of the type described in \cite{r}.

It was argued in \cite{suss,peet}
that a regular configuration cannot dynamically evolve into
a singular one. We comment on how the scenario presented here
evades that argument.
Furthermore, an application of this scenario in the
$1 + 1$ dimensional context described in \cite{suss} suggests that
a physical black hole can dynamically evolve into a massive
remnant described by the static, singular solution of \cite{suss,num}.
This would make massive remnant a likely candidate for DBH.

In this paper we first describe the
static solutions of $d = 2$ string and present a
scenario for dynamical evolution when a localised tachyon pulse
is incident. We then comment on the arguments of \cite{peet},
apply our scenario to the $1 + 1$ dimensional
case of \cite{suss} and conclude with a brief summary.

\vspace{4ex}

\noindent II.
We now describe the static solutions of $d = 2$ string. See \cite{r}
for details. The $\beta$-function equations
for graviton $(G_{\mu \nu})$, dilaton $(\phi)$, and tachyon $(T)$
are
\begin{eqnarray}\label{beta}
R_{\mu \nu} + \nabla_{\mu} \nabla_{\nu} \phi
+ \nabla_{\mu} T \nabla_{\nu} T & = & 0 \nonumber \\
R + (\nabla \phi)^2 + 2 \nabla^2 \phi + (\nabla T)^2
+ 4 \gamma K & = & 0 \nonumber \\
\nabla^2 T + \nabla \phi \nabla T - 2 \gamma K_T & = & 0
\end{eqnarray}
where
$ \gamma = \frac{- 2}{\alpha'}, \;
K = 1 + \frac{V}{4 \gamma} , \;
V = \gamma T^2 + {\cal O} (T^3)$ is the tachyon potential
and $K_T \equiv \frac{d K}{d T}$.
These equations also follow from the effective action
\begin{equation}\label{target}
S = \int d^2 x \sqrt{G} \, e^{\phi} \, ( R - (\nabla \phi)^2
+ (\nabla T)^2 + 4 \gamma K )  \nonumber
\end{equation}
in the target space with coordinates
$ x^{\mu}, \; \mu = 0, \, 1$.
As can be seen from equation (\ref{target}), the dilaton field
$e^{- \frac{\phi}{2}}$ acts as a string coupling.
Consider now the
target space conformal gauge $ d s^2 = e^{\sigma} d u \, d v $
where $u = x^0 + x^1$ and $v = x^0 - x^1$. In this gauge
equations (\ref{beta}) become
\begin{eqnarray}\label{betauv}
\partial_u^2 \phi - \partial_u \sigma \partial_u \phi
+ ( \partial_u T )^2  & = & 0 \nonumber \\
\partial_v^2 \phi - \partial_v \sigma \partial_v \phi
+ ( \partial_v T )^2 & = & 0 \nonumber \\
\partial_u \partial_v \sigma + \partial_u \partial_v \phi
+ \partial_u T \partial_v T & = & 0 \nonumber \\
\partial_u \partial_v \phi + \partial_u \phi \partial_v \phi
+ \gamma K e^{\sigma} & = & 0 \nonumber \\
2 \partial_u \partial_v T + \partial_u \phi \partial_v T
+ \partial_v \phi \partial_u T - \gamma K_T e^{\sigma} & = & 0 \; .
\end{eqnarray}
We now define new coordinates
$\xi = u v, \; \chi = u/v$,
and consider the static case when the fields are independent of $\chi$.
Defining further
$e^{\Sigma} = - \gamma \xi e^{\sigma}$ and
$( \cdots )_1 = (\xi \frac{d}{d \xi}) ( \cdots )$
equations (\ref{betauv}) can be written as
\begin{eqnarray}\label{salmon}
\Sigma_{1 1} + \phi_1 \Sigma_1 = \Sigma_{1 1} + \phi_{1 1}
+ T_1^2 & = & 0 \nonumber \\
T_{1 1} + \phi_1 T_1 + \frac{1}{2} e^{\Sigma} K_T =
\phi_{1 1} + \phi_1^2 - e^{\Sigma} K & = & 0  \; .
\end{eqnarray}

If $T = 0$, the above equations lead to $d = 2$ string
black hole solution of \cite{rest} without tachyon back reaction.
These equations can also be solved explicitly with $T \neq 0$
if $V = 0$. In this case one solution is trivial where
$T = constant$ and which again describes the
$d = 2$ string black hole as in \cite{rest}.
There also exists another solution
where the tachyon field is non trivial.
This solution thus incorporates non trivially tachyon
back reaction on $d = 2$ string black hole
and exhibits new features: the original black hole horizon is split
into two and the curvature scalar develops new singularities at these
horizons.

This non trivial solution is given by
\begin{eqnarray}\label{whale}
e^{\phi} & = & \beta_0 \tau (\tau^{\delta} - l)^{-1} \nonumber \\
T - T_0 & = & - \sqrt{\delta - 1} \ln\tau   \nonumber \\
e^{\Sigma} & = & l \tau^{- \frac{1}{(1 + \epsilon)}}
\end{eqnarray}
where $\beta_0$ and $T_0$ are constants,
$l = \pm 1$ is the sign of
$\xi, \; \epsilon \geq 0$ is a new parameter and
$\delta \equiv \frac{1 + 2 \epsilon}{1 + \epsilon}$.
The variable $\tau$ is related to $\xi$ by
\begin{equation}\label{tauxi}
\int_0^{\tau} d \tau (\tau^{\delta} - l)^{-1}
= A (1 + \epsilon) \ln(\frac{\alpha}{l \xi})
\end{equation}
where $\alpha$ is a constant.
The curvature scalar $R$ becomes
\begin{equation}\label{dolphin}
R = 4 \gamma (1 + 2 \epsilon)^{-2} \tau^{- \delta}
(\tau^{\delta} - l) (\epsilon \tau^{\delta} + l (1 + \epsilon)) \; .
\end{equation}

As explained in \cite{r}, the full solution is described by
equations (\ref{whale})-(\ref{dolphin}) in two branches.
The branch I with $l = m = 1$ and $1 \leq \tau \leq \infty$ describes
the region $\infty \geq \xi \geq \xi_+ $ and
the branch II with $l = m = - 1$ and $0 \leq \tau \leq \infty$
describes the region $- \alpha \leq \xi \leq - \xi_- $,
where $\xi_{\pm} (> 0)$ are defined by
\[
\int_0^{\infty} d \tau (\tau^{\delta} \mp 1)^{-1}
= A (1 + \epsilon) \ln(\frac{\alpha}{\xi_{\pm}})
\]
and vanish when $\epsilon = 0$.

The features of the above solution are as follows:

(1) $\epsilon = 0$: In this case the $\tau$-integration in equation
(\ref{tauxi}) can be performed and the dilaton and the graviton
fields are
\begin{equation}\label{bh}
e^{\phi} = e^{- \sigma} = - \gamma (\xi + \alpha)
\end{equation}
where $ - \gamma \alpha = \beta_0$ is the black hole mass parameter.
This is the black hole solution
of \cite{rest}. The horizon is at $\xi = 0$. The curvature
scalar $R$  and the string coupling $e^{- \frac{\phi}{2}}$
are singular only at $\xi = - \alpha$.
The tachyon field $T = T_0 ( = 0 $, if  $V \neq 0$)
corresponds to a trivial configuration and does
not back react on graviton-dilaton system. Asymptotically, $R$
and $e^{- \frac{\phi}{2}}$  vanish.

(2) $\epsilon > 0$: The tachyon field is non trivial and the solution
incorporates tachyon back reaction. The horizon now is split into two
located at $\xi = \pm \xi_{\pm}$. The curvature scalar $R$
and the string coupling $e^{- \frac{\phi}{2}}$
are still singular
at $\xi = - \alpha$ but now they develop new singularities
at the horizons, $\xi = \pm \xi_{\pm}$.
Asymptotically, $R$ and $e^{- \frac{\phi}{2}}$  vanish.

(3) These singular features are not the
result of the approximation $V(T) = 0$. Even when $V \neq 0$
these features persist.

(4) As pointed out by Peet et al. in \cite{peet},
when $V \neq 0$ the static $\beta$-function equations (\ref{salmon})
admit a solution which has non trivial tachyon field and which
is regular at the horizon, thus
retaining the features of the $d = 2$ string black hole.
This soultion can be obtained by expanding the various fields in a
Taylor expansion near the horizon.
It reduces to $T = T_0$ when $V = 0$ and has infinite energy
when $V \neq 0$ \cite{peet}.

\vspace{4ex}

\noindent III.
It is now natural to ask
if the singular configuration described above can be formed dynamically,
say, in a way analogous to the formation of two dimensional
black hole by matter shock waves in \cite{cghs}. Ideally one would
like to solve equations (\ref{betauv}) explicitly
and analyse the evolution of an arbitrary tachyon wave incident in the
asymptotic region. However,
equations (\ref{betauv}) are nonlinear and the general
solutions are difficult to obtain.
Their analysis in the asymptotic region,
as in \cite{dl,russo}, is easier but it may not give any clue
about the fields near the horizon.
For example, the asymptotic static solution in
\cite{r,dl} does not reveal any information about the
singular behaviour of the fields near the horizon.

However, in the absence of an analytic solution or numerical
simulations, one can try to understand the dynamic behaviour
qualitatively. In an attempt towards this goal
we will now present a scenario
in which a tachyon is incident on a black hole and a singular
configuration is likely to result.

Equations (\ref{betauv}) can be considered as describing the
evolution of fields in $u$-direction. Thus one can arbitrarily
assign an initial localised distribution of tachyon on a line
$u = u_0$ in the asymptotic region and study its evolution
towards the horizon.

Consider a $d = 2$ string black hole.
The dilaton field $\phi$ decreases monotonically
from the asymptotic region $(u v = \infty)$ to the horizon
$(u v = 0)$. Now let a tachyon be incident on this black
hole from the asymptotic region
with a localised distribution on the line $u = u_0 >> 1$ given by
\begin{equation}\label{deltach}
(\partial_v T)^2 (u_0, v) = a^2 \delta (v - v_0)
\end{equation}
where $a^2$ is a constant.
The tachyon field along $u$-direction is taken to be localised.
This kind of matter distribution is commonly used
in various toy models \cite{cghs}
to form $1 + 1$ dimensional black hole.
In our case, we do not
exactly need a $\delta$-function in (\ref{deltach}).
Any sharply peaked localised distribution will do, as illustrated by
an example after the discussion below.
However, we will first continue with the $\delta$-function.

We can see the effect of this tachyon field
from the second equation in (\ref{betauv}) which implies that,
on the line $u = u_0$,
\[
- \partial_v^2 \phi = a^2 \delta (v - v_0) + \cdots
\]
where $\cdots$ denote the contributions from $\partial_v \sigma$
and $\partial_v \phi$ which are negligible compared to
$(\partial_v T)^2$.
Physically the above equation implies a discontinuity
in $\partial_v \phi$ whereby the dilaton field dips a little deeper
towards the horizon. Thus,
the dilaton field on the line $u = u_0$ has a cusp at $v = v_0$. Now from
the remaining equations in (\ref{betauv}) one can see how this
cusp evolves as one moves towards the horizon
to lower values of $u$.

The tachyon equation in (\ref{betauv}) gives the evolution of
$\partial_v T$ as
\[
- 2 \partial_u \partial_v T
= \partial_u \phi \partial_v T + \cdots
\]
where $\cdots$ denote contributions from
$\partial_v \phi, \; \partial_u T, \; T,$ and $e^{\sigma}$ all of which
are negligible compared to $\partial_v T$.
Since $u$ decreases as one moves towards the horizon it follows that
$- \partial_u$ represents the rate of change as one moves
towards the horizon  and that $\partial_u \phi$ is positive (since
the dilaton field decreases towards the horizon). Thus
taking $\partial_v T$ to be positive, we see from the above equation
that $\partial_v T$ increases towards the horizon. Hence,
$- \partial_v^2 \phi$ also increases. That is, the cusp in the
dilaton field $\phi$, and therefore the discontinuity
in $\partial_v \phi$, increases as one moves towards the horizon.

The fourth equation in (\ref{betauv}) describes
the evolution of $\partial_v \phi$ as
\[
- \partial_u \partial_v \phi =
\partial_u \phi \partial_v \phi
+ \gamma (1 + \frac{T^2}{4}) e^{\sigma}  \; .
\]
For the initial black hole back ground that is under study,
it can be seen easily that the right hand side of the
above equation is negative. This implies that $\partial_v \phi$
decreases as one moves towards the horizon. Combined with the
behaviour of $\partial_v^2 \phi$ explained above, this
indicates that the cusp in the dilaton distribution
will grow without bound
as one evolves towards the horizon using the full equations
(\ref{betauv}). This will also influence the evolution
of graviton and tachyon through their non linear couplings to dilaton
in equation (\ref{betauv}). If there is a static end to this
evolution, it is likely to be described by
the singular static solution discovered in \cite{r} where the curvature
scalar and the dilaton develop new singularities at the (split) horizon.

In the above discussions the $\delta$-function
in (\ref{deltach}) is unnecessary. It is used only to dominate
the contributions of other fields and their derivatives. But
this can be achieved by any smooth localised
tachyon field with large enough $v$-derivatives.
This will smoothen out the cusp in the dilaton and the above
arguments will still go through as we will now illustarte.

Consider a tachyon pulse given by
\begin{equation}\label{pulse}
(\partial_v T)^2 = \frac{a^2 \lambda f(u - u_0) }
{\pi (\lambda^2 + (v - v_0)^2)}
\end{equation}
where $\lambda \to 0$ and $a^2$ denotes the strength of the
pulse\footnote{The tachyon pulse can also be represented
by a normalised gaussian function, centered at $(u_0, v_0)$,
in the limit when its width in the $v$-direction vanishes. However,
the pulse represented by equation (\ref{pulse}) is more convenient for our
purposes.}.
The function $f$ denotes the localistaion in the $u$-direction and can
be given, for example, by
$f(x) = 1$ if $|x| < L$, and $= e^{\frac{1 - x^2}{l}}$ otherwise,
for some convenient choice of $L$ and $l$.
Thus the tachyon pulse is smoothly varying in
$u$-drirection and is localised. Using the second equation in (\ref{betauv})
one gets, after an integration and using the initial black hole
back ground configuration,
\begin{equation}
\partial_v \phi = \frac{u}{u v + \alpha}
- \frac{a^2}{2} (\frac{2}{\pi} \tan^{- 1} \frac{v - v_0}{\lambda} + 1)
 f(u - u_0) + {\cal O}(\lambda) \; .
\end{equation}
In the limit $\lambda \to 0$, the field $\partial_v \phi$ does
indeed develop a discontinuity, proportional to the strength of the
tachyon pulse.

We will now consider the evolution of the tachyon pulse and
$\partial_v \phi$ in the neighbourhood of
$(u, v) = (u_0, v_0)$, as one moves towards horizon; that is, as $u$
decreases. From the expression for
$\partial_u (\partial_v T)$ in (\ref{betauv})
it follows, after an $u$-integration, that
\begin{equation}
(\partial_v T)^2 = \frac{a^2 \lambda \ln^2 (u v + \alpha)}
{\pi (\lambda^2 + (v - v_0)^2)} f(u - u_0) + \cdots
\end{equation}
where the initial black hole back ground has been used and
$\cdots$ denote subleading terms in the limit $\lambda \to 0$.
{}From the above expression it can be seen that, in the neighbourhood of
$(u_0, v_0)$, the strength of the tachyon pulse increases as one
moves towards the horizon; that is, as $u$ decreases. Similarly from
the fourth equation in (\ref{betauv}) it follows that
\begin{equation}
\partial_u (\partial_v \phi) = \frac{\alpha}{(u v + \alpha)^2}
+ \frac{a^2 \lambda \ln^2(\frac{v - v_0}{\lambda}
+ \sqrt{1 + (\frac{v - v_0}{\lambda} )^2})}{4 \pi (u v + \alpha)}
f(u - u_0) + \cdots \; .
\end{equation}
The right hand side of the above equation is positive near
$(u_0, v_0)$, and hence, the field $\partial_v \phi$ decreases as one
moves towards the horizon. Furthermore,
since the tachyon pulse increases in strength as one moves towards the
horizon, the next iteration using the second equation in (\ref{betauv})
shows that the discontinuity in $\partial_v \phi$ also increases
towards the horizon.

Thus one sees that the tachyon pulse and the cusp in the dilaton
field grow stronger as one moves towards the horizon.
Hence as discussed above, if there is a static end
to this evolution of a tachyon pulse thrown into the black hole,
it is very likely
to be the singular static solution described in \cite{r} whose nature
is completely different from that of a black hole.
However, if the
initial tachyon wave is smooth and weak enough without any
rough perturbation in its evolution,
then it may evolve into the smooth configuration
described in \cite{peet}, which is that of a black hole with
non zero tachyon field. Thus it appears that any generic tachyon pulse
(or any irregularity in a typical tachyon wave)
thrown into the black hole destabilises it and turns it into
a singular object with no resemblence to the original black hole.

The scenario and the analysis described here can be likened to placing a
localised charge on a perfect conductor and
analysing its subsequent distribution. Only, here
the tachyon pulse localised at $v = v_0$ and placed on the line $u = u_0$
eventually grows without bound and destabilises the original system
(the $u$-coordinate here is analogous to time, and moving
towards horizon is analogous to evoloving in time).

The example given above describes the behaviour of a tachyon pulse
near $(u_0, v_0)$. The expressions in this example are valid only in the
neghbourhood of $(u_0, v_0)$. But they
illustrate the features of our scenario
which will very likely lead to the formation of a singular object
starting from a non singular one. However, it is desirable to obtain an
explicit analytic or numerical
solution to equations (\ref{betauv}) that describe
this process in full detail and are valid everywhere.
Work on this project is in progress.

\vspace{4ex}

\noindent IV.
Peet et al. argued in \cite{peet} that
a regular configuration cannot evolve into a singular one,
based on the assumption that for the string coupling
$e^{-\frac{\phi}{2}}$ to blow up, it must first develop
a local maximum, accompanied by a local minimum.
They then show that the maximum-minimum can never seperate and hence
no singular configuration can evolve. Their assumption
is not correct. For one thing, a blow up can occur as in our scenario
starting from a cusp which is neither a maximum nor a minimum
and which can be seeded by an incoming tachyon pulse. For another,
suppose that the string coupling diverging at the horizon {\em did}
evolve from a maximum-minimum. Then when the ``maximum''
diverges to $+ \infty$ corresponding to infinite
string coupling at one edge of the
horizon, the accompanying minimum
would diverge to $- \infty$
at presumably the other edge of the horizon
corresponding to vanishing string coupling there.
But this case would not correspond to the singular solution of
\cite{r} where the string coupling diverges
to $+ \infty$ at {\em both} the edges
of the horizon. In contrast, in the scenario proposed here the cusp
in the string coupling would diverge to $+ \infty$, with its
two sides conceivably at the two edges of the horizon, as in \cite{r}.

 Ideally, one would like to solve the full non linear equations
(\ref{betauv}) and understand the evolution of black hole when an
arbitrary tachyon pulse/wave is incident on it. However,
solving these equations analytically is a difficult task. Perhaps
numerical calculations will provide some insight into this process.

\vspace{4ex}

\noindent V.
The scenario described here can be applied in a different
context. In ref. \cite{suss} (see also \cite{num})
the Hawking radiation and its back reaction
were analysed in $1 + 1$ dimensions and
a static singular solution was found in which the curvature scalar
reaches a maximum before diverging to $- \infty$ at the horizon.
Simultaneously, the string coupling also reaches a maximum and
then tends to zero at the horizon.
This solution could represent \cite{suss} a
massive remnant \cite{gidd}, a candidate for DBH.
But using an argument similar to the one
in \cite{peet} the authors of \cite{suss} argue that
the string coupling in the case of
a $1 + 1$ dimensional black hole \cite{cghs}
cannot develop a maximum before the horizon and hence
conclude that the black holes formed in a collapse process cannot
evolve into a massive remnant described by the singular solution
in \cite{suss,num}.

However, in a scenario similar to the one proposed here,
the string coupling can develop a cusp
if the incident matter distribution (or any perturbation in it)
is localised and peaked sharply enough. This cusp would evade
the arguments given in \cite{suss,peet},
and can evolve into a maximum before the horizon, leading
very likely to the singular configuration of \cite{suss}
which could well be a massive remnant,
thus strengthening its case as a DBH.

Though these remnants do not come with no strings attached, so to
say, they could potentially describe DBH,
at least in $1 + 1$ dimensions
\cite{suss}-\cite{gidd} --- they can carry
enough information in them to solve the ``information puzzle''
and do not have problems associated with
infinite density of states that plague the planck mass remnants.
The massive remnants have problems associated with
causality \cite{gidd} and
naked singularity (if described by the static solutions of
\cite{suss}). Also, understanding its spectrum and the physics
responsible for stopping the Hawking radiation, storing the
information, etc. \ are challenging problems.
If the massive remnants have curvature singularities
as in \cite{suss}, then semiclassical theories may not be applicable
in the regions of strong curvature and the objections
raised by Preskill in \cite{info} against massive remnants may have to
wait until one understands how to deal with the (naked) singularities.
See \cite{cghs}-\cite{info} and references therein
for some of the recent reviews which deal with the issues related to
the massive remnants.

\vspace{4ex}

\noindent VI.
To summarise, the $d = 2$ string and the $1 + 1$ dimensional
toy models appear to be rich enough to describe
the formation and the evolution of black holes. It also appears possible
that the end state of a black hole could be described by a static
singular solution and that the
equations in the above models
could very well describe such an evolution process.
But these equations are nonlinear and hard to solve.
What we reported here is a possible scenario of black hole evolution
according to these equations. It would be desirable
and worthwhile to understand the evolution process as fully as
possible, perhaps with the help of numerical computations.

\vspace{3ex}

It is a pleasure to thank S. Sen for encouragement and to acknowledge
a helpful suggestion from the referee.
This work is supported by EOLAS Scientific Research Program
SC/92/206.

\end{document}